\documentclass[a4paper,twocolumn]{article}
\usepackage[T1]{fontenc}
\usepackage{graphicx}
\usepackage{amsmath}
\usepackage{bm}
\usepackage{amsfonts}
\usepackage{amssymb}
\usepackage{physics}
\usepackage[separate-uncertainty=true]{siunitx}
\usepackage{hyperref}
\usepackage{braket}
\usepackage{placeins}
\usepackage{color}
\usepackage{colordvi}
\usepackage{authblk}
\usepackage{fullpage}

\usepackage{cite}
\begin{document}

\title{Transfer Matrix Model for Emission Profile Optimization of Radial Gratings}

\author[1]{Stefan Appel}
\author[2]{Viviana Villafane}
\author[2]{Jonathan J. Finley}
\author[1,*]{Kai M\"uller}
\affil[1]{Walter Schottky Institut, Department of Electrical and Computer Engineering and MCQST, Technische Universit\"at M\"unchen, 85748 Garching, Germany}
\affil[2]{Walter Schottky Institut, Physics Department, School of Natural Sciences and MCQST, Technische Universit\"at M\"unchen, 85748 Garching, Germany}
\affil[*]{\url{kai.muller@wsi.tum.de}}

\date{\today}

\maketitle

\begin{abstract}
Radial Bragg gratings are commonly used to enhance light extraction from quantum emitters, but lack a well-suited, fast simulation method for optimization beyond periodic designs.
To overcome this limitation, we propose and demonstrate an algorithm based on the transfer matrix model (TMM) to calculate the free-space emission of such gratings.
Using finite difference time domain (FDTD) simulations, we characterize the free-space emission and transfer matrices of single grating components. Our TMM then combines any number of components to receive the total emission. Randomized benchmarks verify that results from our method agree within \SI{98}{\percent} with FDTD while reducing simulation time by one to two orders of magnitude.
The speed advantage of our approach is shown by maximizing emission of a fifteen-trench circular grating into a Gaussian mode.
We expect that our novel algorithm will facilitate the optimization of radial gratings, enabling quantum light sources with unprecedented collection efficiencies.
\end{abstract}

\section{Introduction}
Radial symmetric structures such as radial Bragg gratings or Bullseye resonators are of great interest to enhance collection efficiency and emission rate of single quantum emitters. Most designs thereby employ a periodic grating design \cite{Davanco2011,Ates2012,Li2015b,Winkler2018,Yao2018,Duong2018,Rickert2019,Kolatschek2019}.

Transfer matrix models (TMMs) have been used for decades to predict the behaviour of flat, stratified optical systems, such as distributed Bragg reflectors (DBRs), subject to plane wave illumination \cite{Weinstein1947,Epstein1952,Abeles1957}.
Already in the 1990s, TMMs were adapted for radially curved layers with impinging radial propagating waves \cite{Jiang1994,Ping1994,Kaliteevskii1999}. This also led to improved designs for radial DBRs, deviating from the strict periodicity of planar DBRs to compensate for effects only occurring in radial wave propagation \cite{Kaliteevskii1999,Ochoa2000,Tobar2005,Ben-Bassat2015}. 
In the non-curved case, it was also shown that non-periodic linear grating couplers can scatter light into a target mode more efficiently than periodic ones \cite{Taillaert2006,Lu2007,Zhang2013}.

Even though both the curved geometry and the results for linear waveguide couplers therefore suggest a non-periodic design for radial Bragg gratings, most publications today still revert to use a plane wave TMM and/or periodic radial structure \cite{Davanco2011,Ates2012,Li2015b,Winkler2018,Yao2018,Duong2018,Rickert2019,Kolatschek2019}.
The large computational cost of simulating any radial structure using the most common, rectangular, three-dimensional finite difference time domain (FDTD) method hinders the development of non-periodic radial geometries, and thus, their implementation in optical nanostructures. To make optimization feasible, a periodic design is preferred with a small number of free parameters. This reduces the number of design iterations needed for optimization, such that the total computational requirements remain manageable.

For an alternative simulation method, Li et.al. \cite{Li2021} introduced the idea to use TMM to predict the field scattered into free-space from structures such as non-periodic gratings. Their two-dimensional method is fast and shows excellent overlap with two-dimensional FDTD simulation. Their TMM however is limited to non-curved geometries, even though they used its results as a starting point to design a non-periodic radial grating.
To overcome the computational hurdles associated with non-periodic radial grating designs and expand on the work of Li et.al. \cite{Li2021} we present here a radial scatter-field TMM (rsTMM), capable of describing the free-space emission of radial symmetric shells of scatterers in a single- or few-mode supporting slab waveguide. The performance of the rsTMM is confirmed by comparing results of simulations to two-dimensional radial FDTD (rFDTD) \cite{Oskooi2010} for various structures. Here, an error of less than \SI{2}{\percent} is achieved while saving a factor of 10 to 100 in computation time.

\section{Method Definition}
\begin{figure}[htb]
    \centering
    \includegraphics[width=7cm]{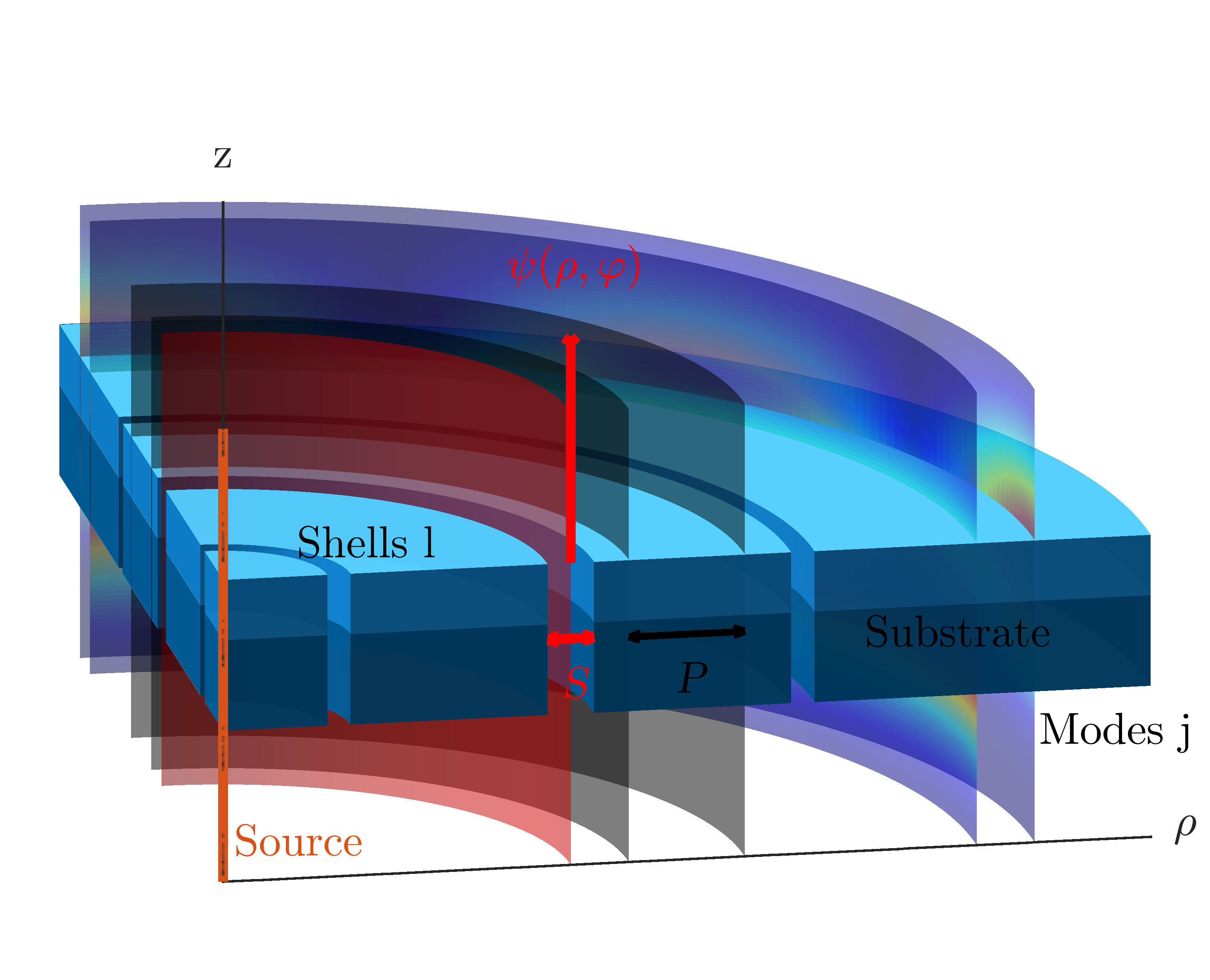}
    \caption{Sketch of the structure to be simulated with the radial scatter transfer matrix model. A centered source is embedded into a layered substrate, supporting few modes $j$. Out- and inward mode propagation within the substrate is described by propagation matrices $P$. Concentric shells $l$ of scatterers surround the source. Their inter-mode scattering is described by scatter matrices $S$, and their mode-dependant emission into free-space by function $\mathbf{\psi}(\rho,\varphi)$.}
    \label{fig:stub}
\end{figure}
Our rsTMM aims to describe a structure as presented in figure \ref{fig:stub}. In a planar substrate with a centered source, $M$ circular shells of scatterers are placed, where both the radius $r_l$ and parameters or shape $q_l$ of any scatter layer $l$ may be changed independently within a set of allowed radii $R$ and parameters $Q$. Such devices are commonly known as a Bullseye resonators respectively as a circular, radial or cylindrical DBRs or gratings \cite{Ochoa2000,Tobar2005,Davanco2011,Li2015b,Li2021}.
To explore the parameter space of the complete device, e.g. for optimization, the computational effort will scale unfavorably according to $|R|^M \cdot |Q|^M$.
As such, we describe this system using a TMM instead and will demonstrate this approach to be quick in assembling any combination of scattering shells. Most of the computational effort is then only needed to characterize a single shell in its 2-dimensional parameter space, a computation effort scaling as $|R| \cdot |Q|$. Our results show that when iterating through many different structures with many shell parameters to vary, our TMM method is faster than directly simulating the structures. Importantly, this approach is not only valid for radial structures as described here, but any kind of structure with many parts that can be separated in a TMM.

As usual for a TMM, individual scatterers as well as mode propagation paths between two scatterers are expressed by transfer matrices $T$, which relate mode amplitudes $a, a'$ at different radial positions $r<r'$, or on the inside and outside of a thin scatterer where $r = r'$ \cite{Seguinot1998,Schmid2015}
\begin{equation}
	a' = T \cdot a
	\label{equ:TMM} \; .
\end{equation}
Both $a$ and $a'$ thereby are $2N$-dimensional vectors, where $N$ is the total number of different modes $j$ considered. One half of entries represent outward ($+$) and the other inward ($-$) traveling mode amplitudes. We determine $T$ for any object by running a set of simulations $i$, recording $a_i$ and $a'_i$ for each. If there is a vector base for the mode amplitude space among $a_i$, then we can calculate $T$ by representing the according amplitude vectors by matrices
\begin{equation}
	A = \left[a_1 \dots a_i \dots a_{2N}\right], A' = \left[a'_1 \dots a'_i \dots a'_{2N}\right]
\end{equation}
and inverting the transfer matrix equation (\ref{equ:TMM}) 
\begin{equation}
	T = A' \cdot A^{-1}
\end{equation}
where the existence of the matrix inverse $A^{-1}$ is guaranteed by the vectors $a_i$ spanning a complete basis.
In practice, we find such a basis by running $2N$ simulations, where for every simulation $i$, we inject only one mode $j$, once on the inside and once on the outside of the object of interest. As inward and outward propagation and scattering behaviour is different due to the curved geometry \cite{Jiang1994,Ping1994}, one cannot assume symmetry to reduce the number of simulations by half \cite{Li2021} (see supplemental material, for further details on the simulation methods).

The modes we consider are guided modes of the planar waveguide substrate at a single vacuum wavelength $\lambda$ of the electro-magnetic field. Substrate examples may be single freestanding slabs or stratified dielectrics, possibly also including metals or dispersive media. All the following expressions depend on $\lambda$ which we omit for clarity. Since we exclude non-linear effects, the vacuum wavelength will not change by propagation or scattering.\\
As the structure of interest is radially symmetric, each guided mode decomposes further into cylindrical modes with symmetry number $m$ \cite{Ludwig1991}. The symmetry number defines the azimuthal field dependence $\propto e^{im\varphi}$ with $\varphi$ the azimuth angle. In total, we collect $N$ individual propagating modes $j$ each characterized by their vertical mode profile and their azimuth propagation symmetry  \cite{Ludwig1991,Cha-MeiTang1979}.

Since we are only interested in electric dipole-like excitation and expect no scattering into modes having higher azimuthal symmetry indices due to the continuous radial symmetry of the structure, we can limit ourselves to $m=0$ for vertical dipoles exciting $E_z$ modes and $m=\pm1$ for in-plane dipoles exciting $H_z$ modes \cite{Cha-MeiTang1979}.
Furthermore, without loosing generality we can combine $m=\pm 1$ to a $\cos(\varphi)$ dependence to represent the symmetry of an in-plane dipole oriented along $\varphi=\pi/2$ \cite{Cha-MeiTang1979}. Note that structures with discrete radial symmetries may be simulated if modal symmetry $m$ is not limited.

To describe the propagation of radial waves in the substrate between two radii $r,r'$, we require the propagation transfer matrices $P(r,r')$. To obtain them, one set of simulations is run without any scatterers present in the substrate, recording the mode amplitudes $a(r)$ at regular spacing. Following equation (\ref{equ:TMM}), we then calculate propagation matrices between arbitrary positions
\begin{equation}
	P(r,r') = A(r') \cdot A(r)^{-1} \; .
\end{equation}
Close to a source at $r=0$, we receive non-zero inward propagating mode amplitudes due to the reactive near field known from antenna theory \cite{Derat2019}. This effect does not appear in flat geometries, since the corresponding plane wave sources are infinitely extended. Also, we observe a non-linear phase evolution close to $r=0$ due to the wavefront curvature. 
These propagation simulations are also used to detect undesired stray fields $\mathbf{\tilde{f'}}_i(\rho,\varphi)$ emitted by imperfect sources into free-space. \\
Similar, in order to describe a single scatter shell $p$ at radius $r$, we compute a transfer matrix $S(r,p)$ by running a set of simulations with the respective shell, recording the mode amplitudes inside $a(r_\text{in})$ and outside $a(r_\text{out})$ of the shell, with $r_\text{in}<r<r_\text{out}$. The total transfer matrix $T$ describing both the scatter as well as the propagation to and from it is then obtained from equation (\ref{equ:TMM}):
\begin{align}
    T(r,q,r_\text{out},r_\text{in}) &= A(r_\text{out}) \cdot A(r_\text{in})^{-1} \nonumber\\
    &= P(r,r_\text{out}) \cdot S(r,q) \cdot P(r_\text{in},r)
\end{align}
Left and right multiplying the inverse of the according propagation matrices $P$, we can then extract $S(r,q)$.

In addition, each shell may emit into free-space when a propagating mode is incident upon it, such that the total emitted fields are given by the coherent sum of the fields emitted by all shells combined \cite{Li2021}. As the field emitted per shell is dependent on which mode was incident from which side, we represent it by a position- and parameter-dependent $2N$-element vector $\boldsymbol{\psi}(r,q,\rho,\varphi)$, with every entry $\psi_j^\pm(r,q,\rho,\varphi)$ representing all electric $\mathbf{E}$ and magnetic $\mathbf{H}$ field components. One half of entries thereby give the fields emitted for a mode impinging from inside outwards ($+$), the other for a mode impinging from outside inwards ($-$).
To recover these free-space fields, for each of the simulations $i$ used to determine $S(r,q)$ before, we record the field $\mathbf{f'}_i(r,q,\rho,\varphi)$ at a fixed distance $z$ above the substrate \cite{Li2021}. For continuous rotational symmetric scatterers, the azimuthal symmetry will be the same as the symmetry of the impinging mode, such that the field may be restored from a record at a single azimuthal angle $\varphi$ only. The previously determined stray fields are then subtracted from imperfect sources to end up with $\mathbf{f}_i(r,q,\rho,\varphi) = \mathbf{f'}_i(r,q,\rho,\varphi) - \mathbf{\tilde{f'}}_i(\rho,\varphi)$. However, the field retrieved in this way may still be created by a multitude of modes impinging on the scatterer from inside and outside, such that
\begin{equation}
    \mathbf{f}_i(r,q,\rho,\varphi) = b_i \cdot \boldsymbol{\psi}(r,q,\rho,\varphi)
    \label{equ:NF}
\end{equation}
with $b_i$ the impinging mode amplitude vector defined as
\begin{equation}
    b_i = \left[a_i^+(r), a_i'^-(r)\right]
    \label{equ:impinge} \; .
\end{equation}
To obtain $\boldsymbol{\psi}(r,q,\rho,\varphi)$, we first calculate the mode amplitudes $a_i(r)$, $a'_i(r)$ on the inside respectively outside of the shell using the propagation matrices
\begin{align}
    a_i(r) &= P(r_\text{in},r) a_i(r_\text{in})\nonumber \\
    a'_i(r) &= P(r,r_\text{out})^{-1} a_i(r_\text{out})
\end{align}
to extract the impinging amplitudes $b_i$. Stacking the results from the set of simulations $i$ to build a matrix $B$ from $b_i$ and a position-dependant vector $\mathbf{F}(\rho,\varphi)$ from $\mathbf{f}_i(\rho,\varphi)$, we can invert equation (\ref{equ:NF}) to calculate
\begin{equation}
    \boldsymbol{\psi}(r,q,\rho,\varphi) = B^{-1} \cdot \mathbf{F}(r,q,\rho,\varphi) \; .
\end{equation}
By repeating this procedure for any number of different scatterers $p$ at different radii $r$, we build a look-up table for transfer matrices $S(r,p)$ and free-space fields $\boldsymbol{\psi}(r,p,\rho,\varphi)$.
Linear interpolation allows us to also retrieve results in between the sampled radii.\\

Using both propagation- and scattering-type transfer matrices, we can now assemble any combination of scatterers at various positions into a multi-shell structure. The according total transfer matrix $T_\text{tot}$ is the product of the alternating scatter and propagation matrices
\begin{align}
    a(r_\text{max}) &= T_\text{tot} \cdot a(r_\text{min}) \nonumber \\
    = &P(r_{M},r_\text{max}) \cdot S(r_M,q_M) \cdot P(r_{M-1},r_{M}) \cdots \nonumber \\
    &P(r_1,r_2) \cdot S(r_1,q_1) \cdot P(r_\text{min},r_1) \cdot a(r_\text{min})
    \label{equ:totTMM} \; .
\end{align}
The total transfer matrix therefore relates mode amplitudes on the inside $a(r_\text{min})$ and outside $a(r_\text{max})$ of the structure.
To fix the actual mode amplitudes, we employ open or impedance-matched boundary conditions on the outside
\begin{align}
    a^+_j(r_\text{max}) = l_j,
    &&a^-_j(r_\text{max}) = 0
\end{align}
with the per-mode side losses $l_j$ remaining as free parameters.\\
On the inside, we inject only outward propagating modes representing the source, initially with zero reflection:
\begin{align}
    \tilde{a}^+_j(r_\text{min}) = s_j,
    &&\tilde{a}^-_j(r_\text{min}) = 0
\end{align}
The complex source amplitudes $s_j$ may be determined by another simulation recording mode amplitudes at $r_{\text{min}}$ with only the desired source present. Note that a multi-mode source may be used, however, the amplitude ratio between the modes may be changed by Purcell effect which will be addressed in future work.

So far we have not addressed inward reflections from the structure, and the boundary conditions $\tilde{a}(r_\text{min})$, $a(r_\text{max})$ cannot be satisfied except for the trivial, scatter-free case. To account for reflections, we have to consider that any reflected, inward propagating mode amplitude, at the center $r=0$, will change into an outward propagating mode amplitude, which then is subject to reflections again. Therefore, we sum up all infinite reflections in a Neumann series and end up with an inner boundary condition
\begin{equation}
    a(r_\text{min}) = \sum_{n=0}^\infty C^n \cdot \tilde{a}(r_\text{min})= \left(I-C\right)^{-1} \cdot \tilde{a}(r_\text{min})
\end{equation}
with the identity $I$ and the center reflection matrix $C$
\begin{align}
    C_{j+,k+} &= 0, \qquad C_{j+,k-} = \delta_{j,k} \varphi_j,  \qquad C_{j-,k-} = 0 \nonumber \\
    C_{j-,k+} &= \sum_l\left(T_\text{tot}^{-1}\right)_{j-,l+} \cdot \left(T_\text{tot}\right)_{l+,k+}. 
\end{align}
$C_{j-,k+}$ describes the reflection of outward-traveling mode $k+$ into inward-traveling mode $j-$ by the structure, $C_{j+,k-}$ the transmission through the center with some phase shift $\varphi_j$ depending on the mode $j$, which would vanish $\varphi_j \to 1$ for $r_\text{min} \to 0$. We determine the phase shift as $\varphi_j = a_j^+(r_\text{min})/a_j^-(r_\text{min})$ using one of the previous simulations with mode injection from the outside.\\
Now that we have taken care of all reflections, the boundary conditions can be fulfilled and we solve for the side-losses $l_j$ using equation (\ref{equ:totTMM}).

Finally, we can now calculate the scattered free-space field of the structure. For this, we first calculate the mode amplitudes inside ($a(r_i)$) and outside ($a'(r_i)$)  of all scattering shells $i$ by successively applying equation (\ref{equ:TMM}), alternating between $P$ and $S$ matrices and starting from the injected amplitudes $a(r_\text{min})$. Using equation (\ref{equ:impinge}), we can then retrieve the impinging amplitudes $b(r_i)$ and from there calculate the emitted fields $\mathbf{f}(r_i,p_i,\rho)$ for every shell by equation (\ref{equ:NF}). Coherently adding up all fields, we obtain the total free-space field
\begin{equation}
    \mathbf{f}_{tot}(\rho,\varphi) = \sum_{i=1}^M \mathbf{f}(r_i,q_i,\rho,\varphi) \; .
\end{equation}

\section{Benchmark Results and Discussion}
To verify our such-defined rsTMM, we simulate a benchmark system and compare the results to established rFDTD \cite{Oskooi2010}.
The benchmark substrate consists of a freestanding, dielectric diamond membrane with refractive index of \SI{2.4114}{} and thickness of \SI{140}{nm}, surrounded by air. It is fully penetrated by rectangular cross-section trenches of varying width and radii. The wavelength of interest is \SI{620}{nm}, leading to one guided $E_z$ and $H_z$ mode each, with effective refractive indices of \SI{1.551}{} and \SI{2.023}{} respectively. Radial symmetry is limited to $m=0$ ($m=1$) for the $E_z$ ($H_z$) mode. For all benchmark situations, centered $E_z$ and $H_z$ sources, coupling to the respective modes, are simulated separately.
We record the free-space field at a distance of $z=\SI{620}{nm}$ above the membrane, and compare it to the results by rFDTD using the normalized squared field overlap \cite{Li2021}
\begin{align}
	&\sigma(\mathbf{f}_\text{rsTMM},\mathbf{f}_\text{rFDTD}) =  \nonumber\\
	&\frac{
		\left|\iint \mathbf{f}_\text{rsTMM}(\rho,\phi) \circ \mathbf{f}_\text{rFDTD}^\ast(\rho,\phi)
		\rho \mathrm{d}\varphi\mathrm{d}\rho\right|^2}
		{
		\iint \left|\mathbf{f}_\text{rsTMM}(\rho,\phi)\right|^2
		\rho \mathrm{d}\varphi\mathrm{d}\rho \cdot 
		\iint \left|\mathbf{f}_\text{rFDTD}(\rho,\phi)\right|^2
		\rho \mathrm{d}\varphi\mathrm{d}\rho
		}  \nonumber \\
		.
\end{align}
For similar results, the overlap should tend towards 1, such that we focus on the difference or error $\Delta \sigma = 1-\sigma$. We separately investigate both the electric $\mathbf{f}=:\mathbf{E}$ and magnetic $\mathbf{f}=:\mathbf{H}$ fields.

\begin{figure}[htb]
    \centering
    \includegraphics[width=\linewidth]{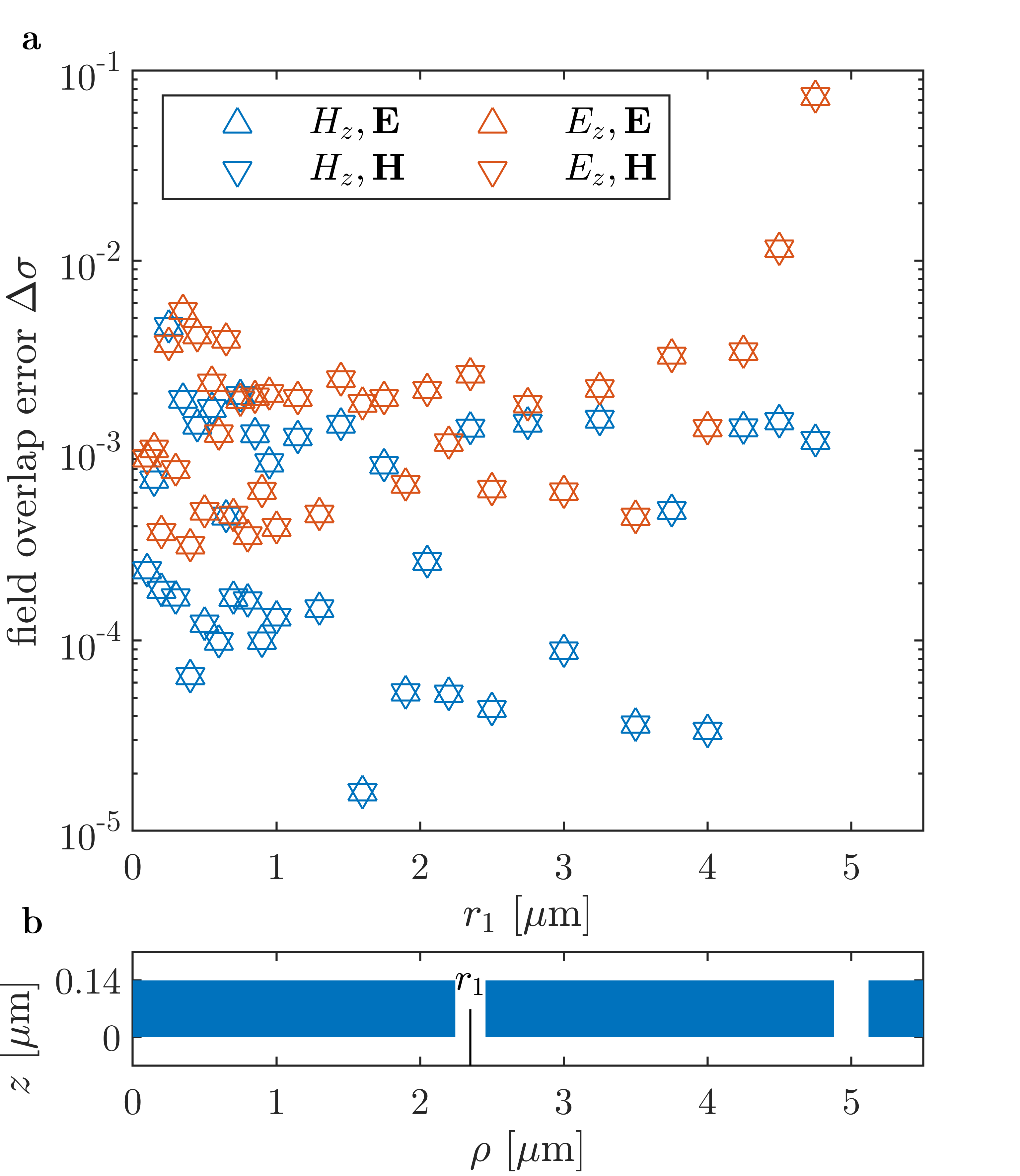}
    \caption{Benchmark of rsTMM for varying radii $r_1$. \\
    \textbf{a} Error in field overlap $\Delta\sigma$ for $\mathbf{E}$  and $\mathbf{H}$ fields between rsTMM and rFDTD for a centered $E_z$ or $H_z$ mode source.\\
    \textbf{b} Example for one of the benchmark structures. The inner trench radius $r_1$ is varied while choosing random trench widths.}
    \label{fig:radiusbench}
\end{figure}

For the first benchmark series, we place two trenches as sketched in figure \ref{fig:radiusbench}b. The inner trench radius is varied continuously while choosing random trench widths, while the outer trench radius is fixed and the trench width is optimized for maximum mode reflection.
Using this arrangement, we can test the amplitude and phase relation between free-space fields created by inner versus outer trench, as well as inward versus outward mode impingement. The error in the field overlap for this series is presented in figure \ref{fig:radiusbench}a, with blue (orange) triangles showing the case for an excitation  with the $H_z$ ($E_z$) mode. Upward (downward) pointing triangles show the error in the $\mathbf{E}$ ($\mathbf{H}$) field respectively.

Importantly, for the data plotted in figure \ref{fig:radiusbench}a the average simulation time was \SI{0.60 \pm 0.03}{s} for rsTMM and \SI{100 \pm 85}{s} for rFDTD, i.e. the rsTMM simulations were more than a factor of 100 faster. For most radii, we find an error of below \SI{1}{\percent} between standard computation techniques and the approach presented in this manuscript. In general, the $H_z$ mode agrees better than the $E_z$ mode.
For large radii of the inner trench, especially the $E_z$ mode shows larger errors. We suspect that this does not stem from the large trench radius but rather from the small distance between the two trenches.
\begin{figure}[htb]
    \centering
    \includegraphics[width=\linewidth]{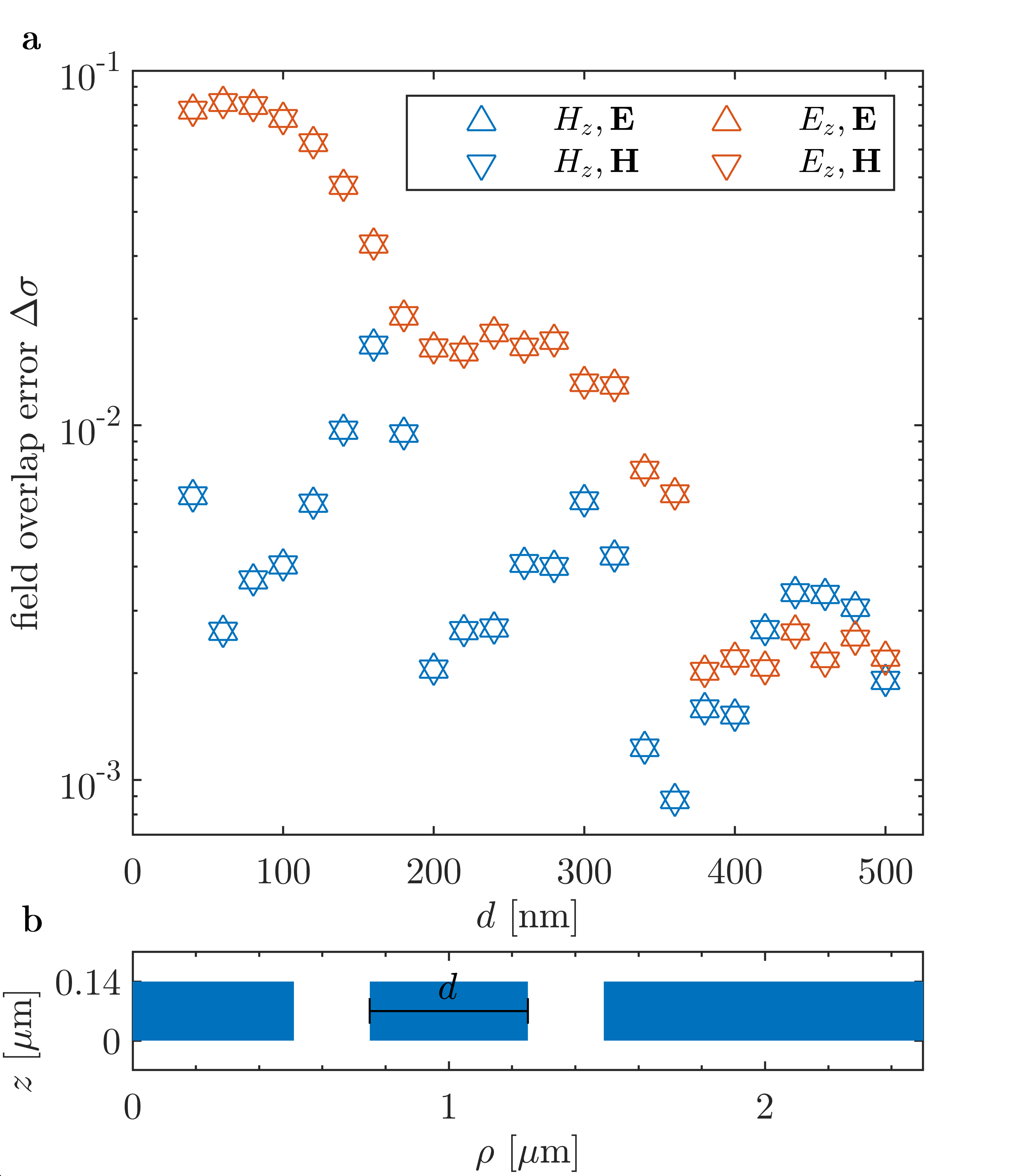}
    \caption{Benchmark of rsTMM for varying trench distances $d$. \\
    \textbf{a} Error in field overlap $\Delta\sigma$ for $\mathbf{E}$  and $\mathbf{H}$ fields between rsTMM and rFDTD for a centered $E_z$ or $H_z$ mode source.\\
    \textbf{b} Example for one of the benchmark structures. The trench distance $d$ is varied with the trench widths fixed.}
    \label{fig:distancebench}
\end{figure}

To investigate this effect with a second benchmark, we place two wide trenches around \SI{1}{\micro m} radius at varying distance to each other as sketched in figure \ref{fig:distancebench}b.
We choose wide trenches to maximize emission into free-space, as we think the effect may be caused by absorption of free-space modes by the neighbouring trench. We do not cover this free-space coupling path with our rsTMM, as we do not include absorption from free-space. As free-space modes should propagate away with distance, we expect a recovery in field overlap with increasing trench distance.
The corresponding error in field overlap for varying trench distances is shown in figure \ref{fig:distancebench}a, again with blue (orange) triangles representing excitation with the $H_z$ ($E_z$) mode and upward (downward) pointing triangles relating to errors in the $\mathbf{E}$ ($\mathbf{H}$) field.

As expected, both modes show increased errors up to \SI{10}{\percent} for trenches placed at small distances, but recover below \SI{1}{\percent} for larger distances.
Superimposed, we observe a beating in the error, which we attribute to resonances in the combined free-space and guided mode coupling.
For sufficiently large radial spacing between scattering elements, our rsTMM method therefore only deviates within \SI{1}{\percent} from the established rFDTD method.
Similar to before, the rsTMM simulations were much faster with simulation times per data point of \SI{0.60 \pm 0.03}{s} for rsTMM and \SI{21\pm10}{s} for rFDTD.

\begin{figure}[!tb]
    \centering
    \includegraphics[width=\linewidth]{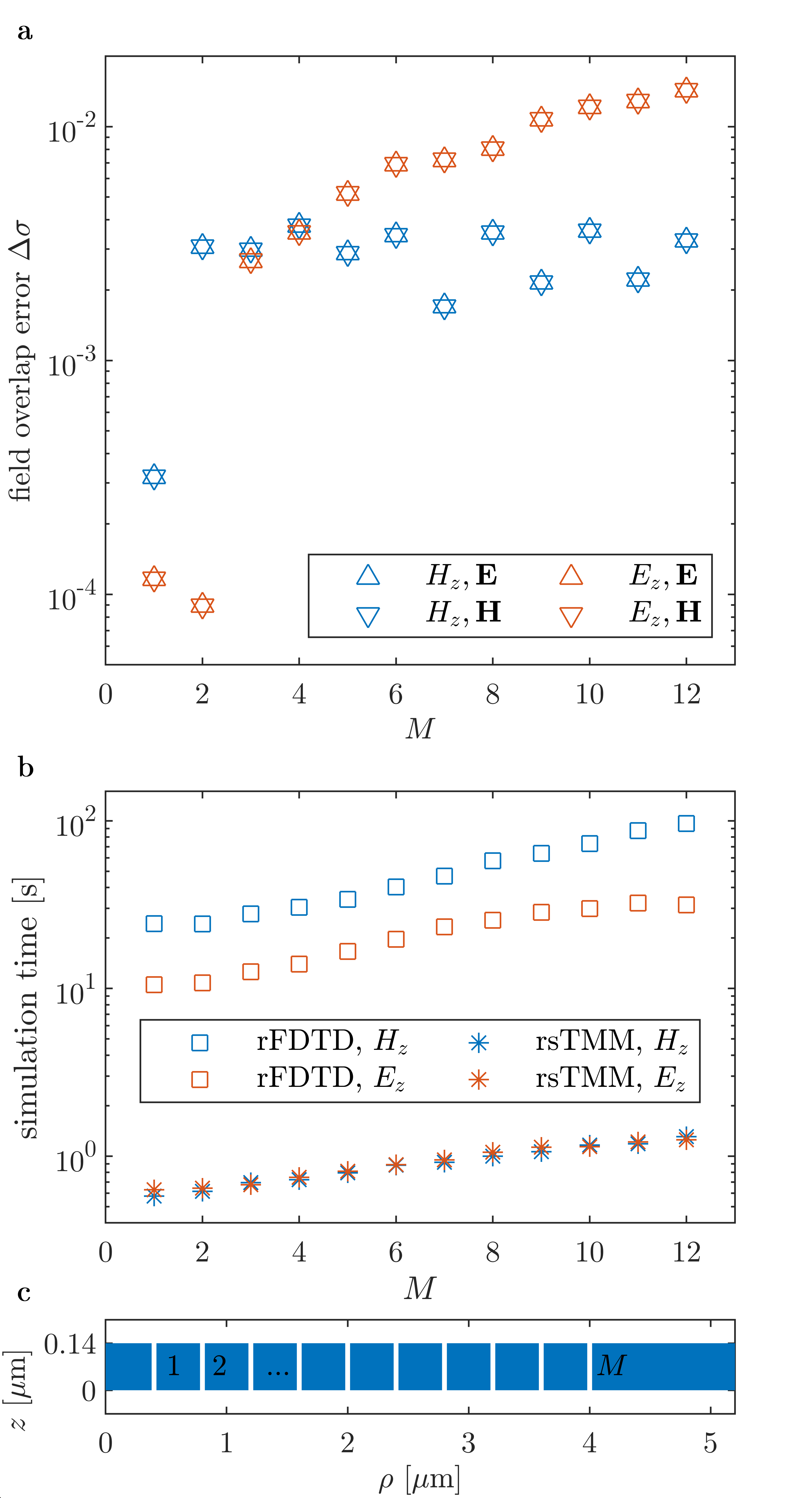}
    \caption{Benchmark of rsTMM for varying trench number $M$. \\
    \textbf{a} Error in field overlap $\Delta\sigma$ for $\mathbf{E}$  and $\mathbf{H}$ fields between rsTMM and rFDTD for a centered $E_z$ or $H_z$ mode source.\\
    \textbf{b} Simulation time required to run the benchmark with rFDTD or rsTMM, for both sources respectively.\\
    \textbf{c} Example for one of the benchmark structures. The number of trenches $M$ is varied with trench widths and distances fixed.}
    \label{fig:numberbench}
\end{figure}
For a final benchmark, we investigate the scaling of computation time and error with increasing number of scattering shells. Thin trenches are chosen to minimize the emission and reflection per trench, such that the contribution of each trench to the total free-space field is comparable. As sketched in figure \ref{fig:numberbench}c, the trenches are placed at a distance of \SI{400}{nm} to each other, in order to minimize the free-space coupling effect, and are added from the center outwards. The error and time requirements are shown in figure \ref{fig:numberbench}a and \ref{fig:numberbench}b respectively. The color scheme follows the previous figures, with blue (orange) representing the excitation with the $H_z$ ($E_z$) mode.

For an increasing number of scatterers, the error between rsTMM and rFDTD  levels off for the $H_z$-mode but keeps increasing for the $E_z$-mode, both for the $\mathbf{E}$ fields (upward triangles) and $\mathbf{H}$ fields (downward triangles). We attribute this to the generally higher error rate in all previous benchmarks for the $E_z$ mode. With an initially higher error for two trenches, the combined error will scale worse with increasing trench number. Importantly, the error still remains within the single-digit percent range.\\
The time for rsTMM (star symbols in figure \ref{fig:numberbench}b) grows approximately linear with the number of trenches, but stays within the few-second regime. We therefore assume that most time in our rsTMM method is spent on interpolations from the scatterer look-up table and possibly on matrix multiplications, while the contribution from the matrix inversion to solve the boundary conditions is small. In contrast, rFDTD simulations on the same machine (open squares in figure \ref{fig:numberbench}b) take between a factor of 10 to 100 longer. The growth law cannot be clearly identified here, but data suggests a significant minimum time requirement. As rFDTD simulations function by propagating fields time step by time step until most of the energy has been emitted from the structure, any extra scatterer reflecting back to the center additionally increases the simulation time.

\section{Optimization Results and Discussion}
\begin{figure*}
    \centering
    \includegraphics[width=\textwidth]{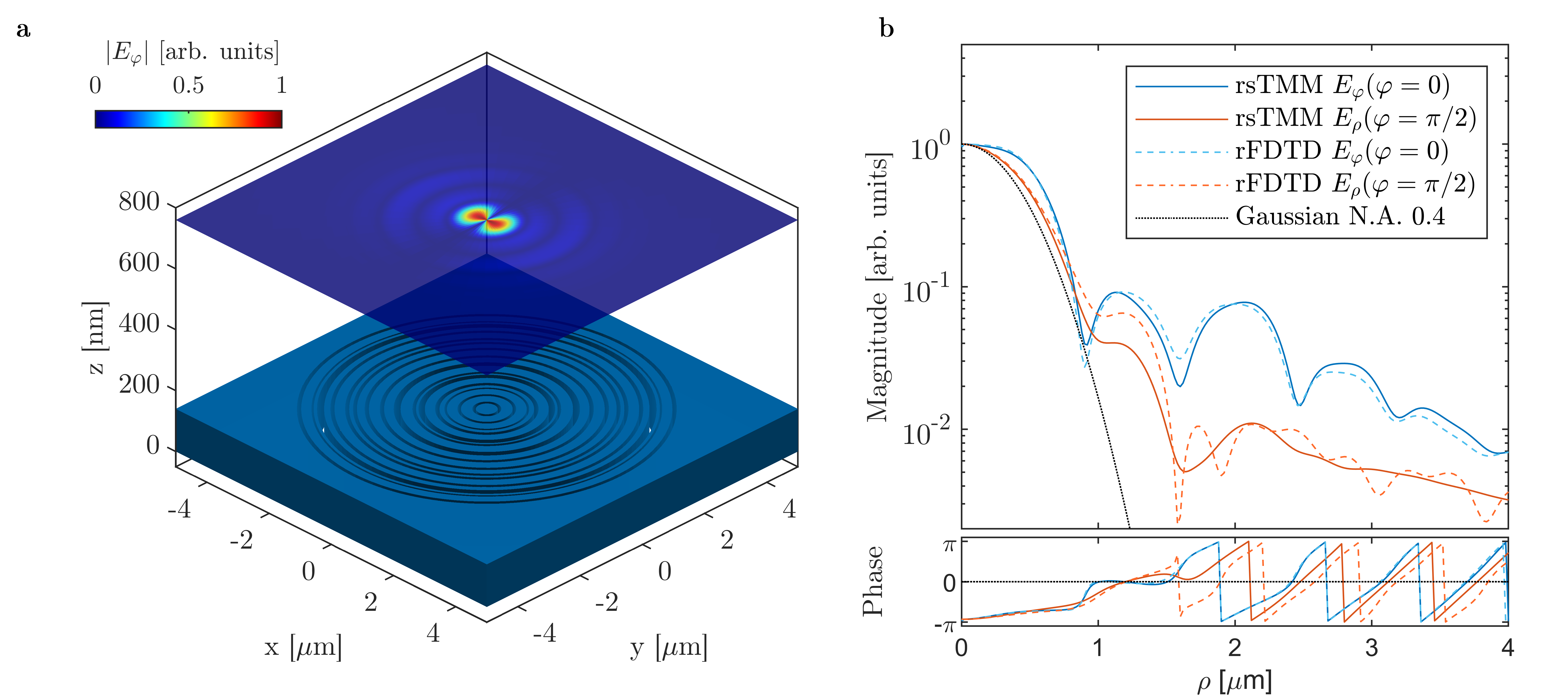}
    \caption{
    \textbf{a} Emission profile (top) achieved with a bullseye structure (bottom), as optimized and simulated using rsTMM.\\
    \textbf{b} Radial field profiles comparing results from rsTMM and rFDTD against the target Gaussian mode.
    The emission into the targeted Gaussian with $N.A. = 0.4$ is \SI{42.44}{\percent}, the overlap between rFDTD and rsTMM \SI{98}{\percent}.}
    \label{fig:optimized}
\end{figure*}
With rsTMM established as much faster but possibly less exact than rFDTD simulations, we now aim to showcase the speed advantage for simulating a large number of structures including lots of scatter shells. To this end, we optimize a Bullseye resonator with $M=15$ fully etched trenches for maximal emission rate from a single mode source centered in the slab into a Gaussian mode, similar as described by Li et.al \cite{Li2021}.
The numerical aperture of the target mode is fixed to \mbox{$NA=0.4$} and both radius $r_i$ and width $w_i$ of each trench are varied individually.
In order to account for fabrication constraints in future experimental realizations, the minimum feature size is set to \SI{50}{nm} and the maximum radius $r_M$ is set to \SI{5}{\micro m}.
Trench distances $\Delta r_i = r_{i+1}-r_i$ are varied between \SI{150}{} and \SI{450}{nm}, and the trench widths are limited to $<\SI{240}{nm}$.
\\
Our figure of merit is the power emitted upward into the target Gaussian mode:
\begin{align}
	FOM = 
	\frac{
	    \mathfrak{Re} \left[\xi_{\text{Target},\text{rsTMM}} \cdot \xi_{\text{rsTMM},\text{Target}} \right] \cdot \Gamma_\text{Top}
		}{
		\left|\xi_{\text{Target},\text{Target}}\right| \cdot \left|\xi_{\text{rsTMM},\text{rsTMM}}\right|
		}
\end{align}
with the overlap integral
\begin{align}
    \xi_{i,j} = \iint
        &\mathbf{E}_i(\rho,\phi) \cross \mathbf{H}_j^\ast(\rho,\phi) \circ \mathbf{\hat{z}}
	~ \rho ~ \mathrm{d}\varphi\mathrm{d}\rho
\end{align}
and the upward emission ratio
\begin{equation}
    \Gamma_\text{Top} = 0.5\cdot\left(1-\frac{\sum_j |a_j^+(r_\text{min})|^2}{ \sum_j |l_j|^2}\right) \; .
\end{equation}
The fraction thereby gives the ratio of power lost through the side, and the pre-factor of 0.5 accounts for the $z$ mirror symmetry, leading to equal top and bottom emission. For the optimization, we use a surrogate algorithm delivered with MATLAB. After simulating 2896 different structures within \SI{1.75}{h} we achieve \SI{42.44}{\percent} top emission into the target Gaussian mode, of \SI{50}{\percent} theoretically possible. The corresponding, optimized structure and its emitted $E_\varphi$-field are shown in figure \ref{fig:optimized}a. Interestingly, the optimized structure (\ref{fig:optimized}a bottom) shows no sign of radial periodicity. To verify our optimization result, we again compare the field profile obtained using rsTMM  and rFDTD simulation. Field profile cuts are plotted in figure \ref{fig:optimized}b as solid (rsTMM) and dashed (rFDTD) lines, along the $x$-axis ($\varphi=0$) for the $E_\varphi$ component (blue) and along the $y$-axis ($\varphi=\pi$) for the $E_\rho$ component (orange). The absolute error between the curves is small; only on the logarithmic scale the relative errors at small field amplitudes become visible. We calculate the error in field overlap between rFDTD and rsTMM to quantify the deviation and find $\Delta \sigma = \SI{2}{\percent}$ both for $\mathbf{E}$ and $\mathbf{H}$ fields. The small increase in error compared to the previous benchmarks results from combining more scattering shells at closer distances.

Finally, we take a look at the time requirements of the two different methods. The simulation time for the final structure was \SI{1.5}{s} for rsTMM and \SI{180}{s} for rFDTD on the same machine.
Assuming a constant simulation time throughout the optimization, we find that the \SI{1.75}{h} total simulation time contains an overhead of \SI{0.5}{h}
from the optimization algorithm. Strikingly, if we had used rFDTD simulations for the optimization, the estimated total optimization time would therefore have been \SI{145}{h}, again assuming a constant simulation time for each structure. For a fair comparison, the time required to create the look-up tables for the scatterers has to be considered as well, which for this benchmark was \SI{63}{h}. This means that even in the worst case, more than \SI{80}{h} of simulation time was saved for one optimization run. However, since the look-up table can be re-used for any further optimization using the same set of scatterers and the same substrate the time advantage per actual use case is significantly higher.

\section{Conclusion}
To summarize, we have introduced a new simulation method for radial symmetric scattering shell structures in planar substrates supporting few guided modes. Our rsTMM agrees within \SI{98}{\percent} with established rFDTD methods except for free-space coupling effects occurring for closely spaced scattering shells. The method is up to 100 times faster than rFDTD, allowing for fast structure optimization in large parameter spaces. We expect that our novel simulation technique will lead to the implementation of more non-periodic and curved nano-photonic grating structures, and may grant access to new collection efficiency records for photonic quantum emitters.
\begin{appendix}
\subsection*{Funding}
We gratefully acknowledge financial support from the German Federal Ministry of Education and Research via the funding program Photonics Research Germany (project MOQUA (13N14846)), the European Union's Horizon 2020 research and innovation program under Grants Agreement No. 862035 (QLUSTER), the Deutsche Forschungsgemeinschaft (DFG, German Research Foundation) via the projects PQET (INST 95/1654-1), MQCL (INST 95/1720-1) and CNLG (MU 4215/4-1), and Germany's Excellence Strategy (MCQST, EXC-2111, 390814868) and the Bavarian State Ministry of Science and Arts via the project EQAP.

\subsection*{Disclosures}
The authors declare no conflicts of interest
\subsection*{Data availability} Data underlying the results presented in this paper may be obtained from the authors upon reasonable request.
\end{appendix}

\bibliography{Collection}

\end{document}


\maketitle

\section*{Radial Finite Difference Time Domain Simulation}
Radial finite difference time domain (rFDTD) simulation is used both to create the look-up table of scatterers as well as for a reference for the benchmarks performed. The simulation set-up is similar in both cases.
We employ the rFDTD algorithm provided with MEEP \cite{Oskooi2010}.
The simulation region spans \SI{7.5}{\mu m} from the origin in radial $\rho$ direction and \SI{1.38}{\mu m} in axial $z$ direction, including symmetric spacers of \SI{620}{nm} above and below the \SI{140}{nm} thick slab. Perfectly matched layers (PML) are added with thickness of \SI{500}{nm} (\SI{400}{nm}) on the $\rho$ ($z$) limits with quadratic profile. Except for the substrate, no other simulation object is extended into the PML. The FDTD cell size is constant and isotropic at \SI{20}{nm}, and all simulations are run until the energy in the simulation volume has decayed beyond $10^{-12}$ of the maximum.

The eigenmodes of the substrate are found by setting up a similar, rectangular simulation in the $x$-$z$ plane and using the included MPB eigenmode solver \cite{Johnson2001} to extract the mode fields $\mathbf{E}_j(z),\mathbf{H}_j(z)$ per mode $j$ for $z$ within the simulation region. TE (TM) modes thereby translate to $H_z$ ($E_z$) modes, and field components $x,y,z$ to radial components $\rho,\varphi,z$. Modes are normalized to satisfy
\begin{align*}
    \frac{1}{2} \int \mathbf{E}_j(z) \cross \mathbf{H}_j^\ast(z) \circ \boldsymbol{\hat{\rho}} ~ dz = 1
\end{align*}
To select the appropriate radial symmetry number $m$, the simulation is set to this symmetry. No scattering between different symmetry modes can be simulated this way, but also does not occur for continuous radial symmetric structures.
Mode sources are set up by placing source regions at $\rho=0$ or $\rho=\SI{6.5}{\mu m}$ spanning $z$, and set to $E_z$ or $H_z$ source type matching the type of mode to excite. The complex amplitudes of the sources are set accordingly to vary with $z$ as $E_{z,j}(z)$ or $H_{z,j}(z)$. The time signal we choose is a Gaussian pulse with a $1/e^2$ spectral width of \SI{100}{nm} centered around $\SI{620}{nm}$, and a cutoff in time at $10^{-8}$ of total amplitude.
To extract mode amplitudes, we place discrete fourier transform (DFT) regions at the desired position $\rho=r_\text{min}\ldots r_\text{max}=\SI{40}{nm}\ldots\SI{5.5}{\mu m}$ spanning $z$ and record all $\mathbf{E}(z)$ and $\mathbf{H}(z)$ field components at the target wavelength(s). To calculate the mode amplitudes, we evaluate the integrals
\begin{align*}
    \alpha_j = \int
    \mathbf{E}(z) \cross
    \mathbf{H}_j^\ast(z) \circ
    \boldsymbol{\hat{\rho}} ~ dz &&
    \beta_j = \int
    \mathbf{E}_j^\ast(z) \cross \mathbf{H}(z) \circ
    \boldsymbol{\hat{\rho}} ~ dz
\end{align*}
and find the inward and outward propagating mode amplitudes as
\begin{align*}
    a^+_j = \frac{\sqrt{\rho}}{4} (\alpha_j+\beta_j) && 
    a^-_j = \frac{\sqrt{\rho}}{4} (\alpha_j-\beta_j)
\end{align*}
with the square root normalizing out the radial field decay.

Scattering into free-space is recorded with a DFT region spanning the top of the simulation region. The fields we record from the simulation thereby relate to $e^{im\varphi}$ symmetry. To receive fields relating to the $\cos(\varphi)$ symmetry of a horizontal dipole, we make the superposition of $m=\pm1$ and employ the symmetry $\{E_\rho,E_\varphi,E_z,H_\rho,H_\varphi,H_z\} \mapsto \{-E_\rho,E_\varphi,-E_z,H_\rho,-H_\varphi,H_z\}$ for $m=+1\mapsto m=-1$, such that we only have to run one simulation at $m=+1$.\\

To prepare the rsTMM, we run sets of simulations with all combinations of inner and outer source positions with both $E_z$ and $H_z$ modes.
For the propagation matrices $P$ and stray source fields $\mathbf{\tilde{f'}}_i(\rho,\varphi)$ we run one such set of simulations with no trenches placed and record mode amplitudes from $\rho=\SI{40}{nm}$ to \SI{5}{\mu m} at \SI{10}{nm} spacing, and additionally at $\rho=\SI{5.5}{\mu m}$.
For the scatterer look-up table, we place the desired trench with width $w$ at radius $r$, but record mode amplitudes only at \SI{40}{nm} and \SI{5.5}{\mu m}. Using one set of simulations each, we record trench widths from \SI{40}{nm} to \SI{240}{nm} at \SI{10}{nm} steps and vary the radius from \SI{100}{nm} to \SI{5}{\mu m} in \SI{100}{nm} steps.
To run rFDTD simulations for comparison in the benchmarks, we keep the same general simulation setup but only employ a mode source at the center and place trenches according to the benchmark design.

\bibliography{Collection}
